# The complete basis with an unextendible product basis and exact-entanglement bases (CBUPB) of three qubit

Xin-Wei[*] Zha, Cun-Bing Huang

(Department of Applied Mathematics and Physics, Xi'An Institute of Posts and Telecommunications,

Xi'an, People's Republic of China 710061)

In this paper, an unextendible product basis and exact-entanglement bases of three qubit is given, and the properties of entanglement for exact-entanglement bases are also discussed. In addition, the bound entangled mixed state is obtained from the exact-entanglement bases.

## I. INTRODUCTION

The concept of bases of a Hilbert space is of paramount importance in quantum mechanics and quantum information. As for multi-particle system in quantum mechanics, basis vector of single-particle pure state is usually used to product, i.e product vector. A product basis (PB) of a Hilbert space of multipartite quantum states is a set of orthogonal product pure states. In common cases, we always use the standard bases, which are the simplest PBs. However, in quantum information, some entanglement basis are likely to be used as base vector. For instance, in the study of teleportation, Bell's bases, which are the maximally entangled states, are very important. Therefore, it is very vital to use some entanglement basis as base vector. Nevertheless, as we known, an arbitrary pure state is either a product state or an entangled state. Therefore, the state of Hilbert space made by an orthogonal complete base may be a product state or an entangled state, as Bell base indicate. However, it is very difficult to find a subspace whose base vector are all entangled and any arbitrary linear combination of the base vector is an entangled pure state. The subspace consisting of the base vector is called exact-entanglement base. [1]

[*] Corresponding Author: Tel: +86-29-85332889, Fax: +86-29-8
 Email:   Zhxw@xiyou.edu.cn (Zha Xin-Wei)

In our work, we get an unextendible product basis （UPB）and exact-entanglement bases of three qubit, and the properties of entanglement for linear combination state of exact-entanglement bases are also discussed. Moreover the bound entangled mixed state is obtained from the exact-entanglement bases.

## II. The EXACT-ENTANGLEMENT BASES OF THREE QUBIT

According to Zai-Zhe Zhong's *Definition*, an entanglement basis (EB) $T = \{|\varphi_0\rangle, \cdots, |\varphi_{n-1}\rangle\}$ is a set of $n$ entangled pure states, $|\varphi_j\rangle (j = 0, \cdots, n-1)$, such that an arbitrary linear combination of them is still an entangled pure state. The subspace $H_T$ spanned by an EB $T$ ($H_T$ does not contain any disentangled pure states) is called an entanglement space (ES). An EB $T$ is called exact-entanglement basis (EEB) if there is a UPB $S = \{|\psi_0\rangle, \cdots, |\psi_{m-1}\rangle\}$ containing $m=N-n$ product states such that $B = S \bigcup T = \{|\psi_0\rangle, \cdots, |\psi_{m-1}\rangle, |\varphi_0\rangle, \cdots |\varphi_{n-1}\rangle\}$ forms an orthogonal complete basis of $H$. In this case, the subspace $H_T$ is called an exact-entanglement space (EES), in which all states and the UPB $S$ are orthogonal to each other. And we call $B$ a complete basis with an

unextendible product basis(CBUPB).

In ref[2], the number of states n in a UPB

$$n \geq \sum_i (d_i - 1) + 1 \tag{1}$$

Therefore, for three qubit, we have $n \geq 4$

For tripartite pure state, there is a UPB $S = \{|\psi_1\rangle, \cdots, |\psi_4\rangle\}$ containing *four* product states; An entanglement basis (EB) $T = \{|\varphi_1\rangle, \cdots, |\varphi_4\rangle\}$ is the *four* entangled pure states, $|\varphi_j\rangle (j = 1, \cdots, 4)$, such that an arbitrary linear combination of them still is an entangled pure state. So that $B = S \cup T = \{|\psi_1\rangle, \cdots, |\psi_4\rangle, |\varphi_1\rangle, \cdots |\varphi_4\rangle\}$ forms an orthogonal complete basis of $H$. And the subspace $H_T$ is called an exact-entanglement space (EES). And we call $B$ a complete basis with an unextendible product basisCBUPB).

As Ref. [3], UPBs for a system of three qubits are

$$|S_1\rangle = |0\rangle \otimes |0\rangle \otimes |0\rangle,$$

$$|S_2\rangle = |1\rangle \otimes |-\rangle \otimes |+\rangle,$$

$$|S_3\rangle = |+\rangle \otimes |1\rangle \otimes |-\rangle, \tag{2}$$

$$|S_4\rangle = |-\rangle \otimes |+\rangle \otimes |1\rangle$$

And $|\pm\rangle = \frac{1}{\sqrt{2}}(|0\rangle \pm |1\rangle)$

Consider a general tripartite pure state written by

$$|\psi\rangle_{ABC} = \sum a_{ijk} |i\rangle_A |j\rangle_B |k\rangle_C \tag{3}$$

Where $i, j, k=0, 1$, the coefficients $a_{ijk}$ s can be arranged as a three-order tensor.

To reduce the size of the expressions, we shall write the components of $|\psi\rangle$ as

$$A_{ijk} = a_r, r = 0, \ldots, 7, \qquad (4)$$

Where r is the integer whose binary expression is ijk, that is, $r = 4i + 2j + k$.

So, any state of three qubit can be written as

$$|\psi\rangle_{123} = a_0|000\rangle + a_1|001\rangle + a_2|010\rangle + a_3|011\rangle + a_4|100\rangle + a_5|101\rangle + a_6|110\rangle + a_7|111\rangle \quad (5)$$

Also satisfying $\langle\psi|\psi\rangle = 1$; **i.e** $\sum_{r=0}^{7}|a_r|^2 = 1.$

So, the UPB of three qubits will be denoted as

$$|S_1\rangle = |000\rangle$$

$$|S_2\rangle = \frac{1}{2}(|100\rangle + |101\rangle - |110\rangle - |111\rangle)$$

$$|S_3\rangle = \frac{1}{2}(|010\rangle - |011\rangle + |110\rangle - |111\rangle),$$

$$|S_4\rangle = \frac{1}{2}(|001\rangle + |011\rangle - |101\rangle - |111\rangle) \qquad (6)$$

According to Ref. [4,5], we can derive

$$\tau_{A(BC)} = 2(1 - tr\rho_A^2)$$
$$= 4\begin{bmatrix}|a_0a_5 - a_1a_4|^2 + |a_0a_6 - a_2a_4|^2 + |a_0a_7 - a_3a_4|^2 \\ +|a_1a_6 - a_2a_5|^2 + |a_1a_7 - a_3a_5|^2 + |a_2a_7 - a_3a_6|^2\end{bmatrix} \qquad (7a)$$

$$\tau_{B(AC)} = 2(1 - tr\rho_B^2)$$
$$= 4\left[\begin{array}{l}|a_0a_3 - a_1a_2|^2 + |a_0a_6 - a_2a_4|^2 + |a_0a_7 - a_2a_5|^2 \\ + |a_1a_6 - a_3a_4|^2 + |a_1a_7 - a_3a_5|^2 + |a_4a_7 - a_5a_6|^2\end{array}\right] \quad (7b)$$

$$\tau_{C(AB)} = 2(1 - tr\rho_C^2)$$
$$= 4\left[\begin{array}{l}|a_0a_5 - a_1a_4|^2 + |a_0a_3 - a_1a_2|^2 + |a_0a_7 - a_1a_6|^2 \\ + |a_3a_4 - a_2a_5|^2 + |a_4a_7 - a_5a_6|^2 + |a_2a_7 - a_3a_6|^2\end{array}\right] \quad (7c)$$

If $\tau_{A(BC)} = 0$, $\tau_{B(AC)} = 0$, $\tau_{C(AB)} = 0$, then the tripartite pure state is fully separable.

It is easy to show for $|S_i\rangle, i = 1,2,3,4$; we have $\tau_{A(BC)} = 0$, $\tau_{B(AC)} = 0$, $\tau_{C(AB)} = 0$.

So for the tripartite pure state, there is a UPB $S = \{|S_1\rangle, \cdots, |S_4\rangle\}$,

For the tripartite pure state, we can give entanglement basis

$$|\varphi_1\rangle = \frac{1}{2}(|001\rangle + |010\rangle + |100\rangle + |111\rangle)$$

$$|\varphi_2\rangle = \frac{1}{2}(|001\rangle - |010\rangle + |101\rangle + |110\rangle)$$

$$|\varphi_3\rangle = \frac{1}{2}(-|001\rangle + |011\rangle + |100\rangle + |110\rangle)$$

$$|\varphi_4\rangle = \frac{1}{2}(|010\rangle + |011\rangle - |100\rangle + |101\rangle) \quad (8)$$

And $\langle \varphi_i | S_j \rangle = 0$

By using Eq. (7)) one obtains, for $|\varphi_j\rangle (j = 1, \cdots, n)$,

$$\tau_{A(BC)} = 1, \tau_{B(AC)} = 1, \tau_{C(AB)} = 1.$$

Furthermore, we find, for $|\varphi_j\rangle (j = 1,\cdots,n)$,

With $\tau_{ABC} = |4H\det(t_{ijk})|$,

Where,

$$H\det(t_{ijk}) = a_0^2 a_7^2 + a_3^2 a_4^2 + a_1^2 a_6^2 + a_2^2 a_5^2 + 4(a_1 a_2 a_4 a_7 + a_0 a_3 a_5 a_6)$$
$$- 2(a_0 a_7 a_1 a_6 + a_0 a_7 a_2 a_5 + a_0 a_7 a_3 a_4 + a_1 a_6 a_2 a_5 + a_1 a_6 a_3 a_4 + a_2 a_5 a_3 a_4)$$
$$= (a_0 a_7 + a_1 a_6 - a_2 a_5 - a_3 a_4)^2 + 4(a_0 a_6 - a_2 a_4)(a_3 a_5 - a_1 a_7)$$

We can obtain $\tau_{ABC} = 1$.

Let $|\psi\rangle = \lambda_1|\varphi_1\rangle + \lambda_2|\varphi_2\rangle + \lambda_3|\varphi_3\rangle + \lambda_4|\varphi_4\rangle$

We can show $|\psi\rangle$ must is entangled state, unless $\lambda_1 = \lambda_2 = \lambda_3 = \lambda_4 = 0$

So that $B = S \cup T = \{S_1,\cdots,|S_4\rangle,|\varphi_1\rangle,\cdots|\varphi_4\rangle\}$ forms an orthogonal complete basis of $H$.

and subspace $H_T$ is called an exact-entanglement space (EES)

## III. THE PROPERTIES OF EES

Let $|\psi\rangle = \lambda(|\varphi_1\rangle + |\varphi_2\rangle)$, then $a_4 = a_5 = a_6 = a_7 = \lambda, a_1 = 2\lambda$;

If $|\psi\rangle$ is satisfy $\langle\psi|\psi\rangle = 1$,

And $|\lambda|^2 = \frac{1}{8}$, we have $\tau_{A(BC)} = \frac{3}{4}, \tau_{B(AC)} = \frac{1}{2}, \tau_{C(AB)} = \frac{1}{2}, \tau_{ABC} = \frac{1}{4}$; using

$$\tau_{AB} = \frac{1}{2}(\tau_{A(BC)} + \tau_{B(AC)} - \tau_{C(AB)} - \tau_{ABC}) \tag{9a}$$

$$\tau_{BC} = \frac{1}{2}\left(\tau_{B(AC)} + \tau_{C(AB)} - \tau_{A(BC)} - \tau_{ABC}\right) \qquad (9b)$$

$$\tau_{AC} = \frac{1}{2}\left(\tau_{A(BC)} + \tau_{C(AB)} - \tau_{B(Ac)} - \tau_{ABC}\right) \qquad (9c)$$

Therefore, $\tau_{AB} = \frac{1}{4}, \tau_{BC} = 0, \tau_{AC} = \frac{1}{4}, \tau_{ABC} = \frac{1}{4}$.

Let $|\psi\rangle = \lambda(|\varphi_1\rangle + |\varphi_2\rangle + |\varphi_3\rangle)$, $|\lambda|^2 = \frac{1}{12}$,

We have $\tau_{A(BC)} = \frac{4}{9}, \tau_{B(AC)} = 0, \tau_{C(AB)} = \frac{4}{9}, \tau_{ABC} = 0$

$\tau_{AB} = \frac{4}{9}, \tau_{BC} = 0, \tau_{AC} = \frac{4}{9}, \tau_{ABC} = 0$

Also, Let $|\psi\rangle = \lambda(|\varphi_1\rangle + |\varphi_2\rangle + |\varphi_3\rangle + |\varphi_4\rangle)$, If $|\psi\rangle$ is satisfy $\langle\psi|\psi\rangle = 1$, and $|\lambda|^2 = \frac{1}{16}$,

we have $\tau_{A(BC)} = \frac{3}{8}, \tau_{B(AC)} = \frac{3}{8}, \tau_{C(AB)} = \frac{3}{8}. \tau_{ABC} = \frac{3}{16};$

$\tau_{AB} = \frac{3}{32}, \tau_{BC} = \frac{3}{32}, \tau_{AC} = \frac{3}{32}, \tau_{ABC} = \frac{3}{16}$.

**IV. DISCUSSION AND CONCLUSION**

Also, we can obtain

$$|S_1\rangle = |111\rangle$$

$$|S_2\rangle = \frac{1}{2}(|011\rangle + |010\rangle - |001\rangle - |000\rangle)$$

$$|S_3\rangle = \frac{1}{2}(|101\rangle - |100\rangle + |001\rangle - |000\rangle),$$

$$|S_4\rangle = \frac{1}{2}(|110\rangle + |100\rangle - |010\rangle - |000\rangle)$$

$$|\varphi_1\rangle = \frac{1}{2}(|110\rangle + |101\rangle + |011\rangle + |000\rangle)$$

$$|\varphi_2\rangle = \frac{1}{2}(|110\rangle - |101\rangle + |010\rangle + |001\rangle)$$

$$|\varphi_3\rangle = \frac{1}{2}(-|110\rangle + |100\rangle + |011\rangle + |001\rangle)$$

$$|\varphi_4\rangle = \frac{1}{2}(|101\rangle + |100\rangle - |011\rangle + |010\rangle) \tag{9}$$

Actually, As ref[1], the orthogonal complete basis $B = S \cup T = \{|S_1\rangle, \cdots, |S_4\rangle, |\varphi_1\rangle, \cdots |\varphi_4\rangle\}$ is invariant under a local operation as

$$B \to B' = S' \cup T' = \{|S'_1\rangle, \cdots, |S'_4\rangle, |\varphi'_1\rangle, \cdots |\varphi'_4\rangle\}$$

$$|S'_i\rangle = u_1 \otimes u_2 \otimes u_3 |S_i\rangle \quad i = 1,2,3,4$$

$$|\varphi'_j\rangle = u_1 \otimes u_2 \otimes u_3 |\varphi_j\rangle \quad j = 1,2,3,4$$

Where, $u_i, i = 1,2,3$ are the arbitrary $2 \times 2$ unitary matrixes.

Therefore, from the above orthogonal complete basis one can create various orthogonal complete bases (in the same type).

If in the system there is an EEB $T = \{|\varphi_1\rangle, \cdots, |\varphi_n\rangle\}$, then the uniform mixture $\tilde{\rho} = \frac{1}{n}\sum_{j=1}^{n}|\varphi_j\rangle\langle\varphi_j|$ is entangled and is a bound entangled state.

Therefore, we obtain the following bound entangled.

$$\tilde{\rho} = \frac{1}{n}\sum_{j=1}^{n}|\varphi_j\rangle\langle\varphi_j| = \frac{1}{4}\sum_{j=1}^{4}|\varphi_j\rangle\langle\varphi_j| = \frac{1}{16}\begin{pmatrix} 0 & 0 & 0 & 0 & 0 & 0 & 0 & 0 \\ o & 3 & 0 & -1 & 0 & 1 & 0 & 1 \\ 0 & 0 & 3 & 1 & 0 & 0 & -1 & 1 \\ 0 & -1 & 1 & 2 & 0 & 1 & 1 & 0 \\ 0 & 0 & 0 & 0 & 3 & -1 & 1 & 1 \\ 0 & 1 & 0 & 1 & -1 & 2 & 1 & 0 \\ 0 & 0 & -1 & 1 & 1 & 1 & 2 & 0 \\ 0 & 1 & 1 & 0 & 1 & 0 & 0 & 1 \end{pmatrix}$$

## ACKNOWLEDGEMENT

The authors gratefully acknowledge the support of Shaanxi Province Natural Science Foundation (2004A15), Science and Technology Plan Foundation of Education of Shaanxi Province (05Jk288).